\documentclass[suppldata]{interact}

\usepackage{epstopdf}
\usepackage[caption=false]{subfig}
\usepackage[T1]{fontenc}
\usepackage{longtable}
\usepackage[numbers,sort&compress]{natbib}
\usepackage{tabularx}
\usepackage{relsize}
\usepackage{url}
\usepackage[hidelinks]{hyperref}
\usepackage{tikz}
\usetikzlibrary{decorations.pathreplacing}
\usetikzlibrary{shapes.geometric}
\usetikzlibrary{shapes.arrows}
\usetikzlibrary{backgrounds}
\usepackage{lmodern}
\usepackage{fancyvrb}
\usepackage{listings}
\definecolor{darkgray}{rgb}{.4,.4,.4}
\lstdefinelanguage{nextflow}[]{Java}{
keywordstyle=\color{black}\fontencoding{T1}\fontfamily{lmtt}\fontseries{b}\selectfont\selectfont,
  stringstyle=\color{black!60!white}\ttfamily,
  keywords=[3]{each, findAll, groupBy, collect, inject, eachWithIndex},
  morekeywords={def, as, in, use, require, promise, process, input},
  moredelim=[is][\textcolor{darkgray}]{\%\%}{\%\%},
  moredelim=[il][\textcolor{darkgray}]{§§}
}
\DeclareSymbolFont{lasy}{U}{lasy}{m}{n}
\SetSymbolFont{lasy}{bold}{U}{lasy}{b}{n}
\DeclareMathSymbol\leadsto{\mathrel}{lasy}{"3B}

\bibpunct[, ]{[}{]}{,}{n}{,}{,}

\theoremstyle{plain}
\newtheorem{theorem}{Theorem}[section]

\theoremstyle{definition}
\newtheorem{definition}[theorem]{Definition}

\theoremstyle{remark}

\definecolor{Gray}{gray}{0.9}

\renewcommand{\phi}{\varphi}
\newcommand{\figref}[1]{\figurename~\ref{#1}}

\newcommand{\tabref}[1]{Table~\ref{#1}}
\newcommand{\secref}[1]{§\ref{#1}}

\newcommand{\NOMAD}{{\smaller NOMAD}}
\newcommand{\FORCE}{{\smaller FORCE}}
\newcommand{\AWDL}{{\smaller AWDL}}
\newcommand{\FHIaims}{{\smaller FHI}-aims}

\newcommand{\PROV}{{\smaller PROV}}
\newcommand{\RAID}{{\smaller RAID}}

\newcommand{\VASP}{{\smaller VASP}}

\newcommand{\API}{{\smaller API}}
\newcommand{\CPU}{{\smaller CPU}}
\newcommand{\CWL}{{\smaller CWL}}
\newcommand{\DAG}{{\smaller DAG}}
\newcommand{\DAW}{{\smaller DAW}}
\newcommand{\DFS}{{\smaller DFS}}
\newcommand{\DFT}{{\smaller DFT}}
\newcommand{\DSL}{{\smaller DSL}}
\newcommand{\GPU}{{\smaller GPU}}
\newcommand{\HDD}{{\smaller HDD}}

\newcommand{\ONE}{{\smaller ONE}}
\newcommand{\RAM}{{\smaller RAM}}
\newcommand{\SSD}{{\smaller SSD}}
\newcommand{\UML}{{\smaller UML}}
\newcommand{\XML}{{\smaller XML}}

\newcommand{\EE}{{\smaller EE}}
\newcommand{\EO}{{\smaller EO}}
\newcommand{\IC}{{\smaller IC}}
\newcommand{\IO}{{\smaller I/O}}
\newcommand{\IT}{{\smaller IT}}
\newcommand{\RM}{{\smaller RM}}
\newcommand{\SQL}{{\smaller SQL}}
\newcommand{\VC}{{\smaller VC}}

\newcommand{\M}{{\smaller M}}
\newcommand{\Ss}{{\smaller S}}

\newcommand{\exciting}{{\usefont{T1}{lmtt}{b}{n}exciting}}

\usepackage{fontawesome5} 
\newcommand{\orcid}[1]{\href{https://orcid.org/#1}{\textcolor[HTML]{A6CE39}{$^{\textrm{\faOrcid}}$}}}

\begin{document}

\articletype{ARTICLE}

\title{Validity Constraints for Data Analysis Workflows}

\newcommand{\ZIB}{\textsuperscript{1}}
\newcommand{\HU}{\textsuperscript{2}}
\newcommand{\trier}{\textsuperscript{3}}
\newcommand{\MDC}{\textsuperscript{4}}
\newcommand{\FHI}{\textsuperscript{5}}

\author{
  \name{Florian~Schintke\ZIB{}\orcid{0000-0003-4548-788X},
    Ninon~De~Mecquenem\HU{}\orcid{0000-0003-3052-6129},
    David~Frantz\trier{}\orcid{0000-0002-9292-3931},
    Vanessa~Emanuela~Guarino\HU{}\textsuperscript{,}\MDC{}\orcid{0000-0003-1625-4323},
    Marcus~Hilbrich\HU{}\orcid{0000-0003-3717-9449},
    Fabian~Lehmann\HU{}\orcid{0000-0003-0520-0792},
    Rebecca~Sattler\HU{}\orcid{0000-0002-7342-7131},
    Jan~Arne~Sparka\HU{}\orcid{0000-0002-5886-4595},
    Daniel~Speckhard\HU{}\textsuperscript{,}\FHI{}\orcid{0000-0002-9849-0022},
    Hermann~Stolte\HU{}\orcid{0000-0003-0047-5842},
    Anh~Duc~Vu\HU{}\orcid{0000-0003-4035-2804},
    Ulf~Leser\HU{}\orcid{0000-0003-2166-9582}}
\affil{
  \ZIB{}Zuse Institute Berlin, Berlin%
  ; \HU{}Humboldt-Universität zu Berlin, Berlin%
  ; \trier{}Trier University%
  ; \MDC{}Max-Delbrück Center for Molecular Medicine, Berlin%
  ; \FHI{}Fritz-Haber-Institut der Max-Planck-Gesellschaft, Berlin%
}}
\date{}
\maketitle

\begin{abstract}
Porting a scientific data analysis workflow (\DAW{}) to a cluster infrastructure, a new software stack, or even only a new dataset with some notably different properties is often challenging. Despite the structured definition of the steps (tasks) and their interdependencies during a complex data analysis in the \DAW{} specification, relevant assumptions may remain unspecified and implicit. Such hidden assumptions often lead to crashing tasks without a reasonable error message, poor performance in general, non-terminating executions, or silent wrong results of the \DAW{}, to name only a few possible consequences. Searching for the causes of such errors and drawbacks in a distributed compute cluster managed by a complex infrastructure stack, where \DAW{}s for large datasets typically are executed, can be tedious and time-consuming.

We propose validity constraints (\VC{}s) as a new concept for \DAW{} languages to alleviate this situation. A \VC{} is a constraint specifying some logical conditions that must be fulfilled at certain times for \DAW{} executions to be valid. When defined together with a \DAW{}, \VC{}s help to improve the portability, adaptability, and reusability of \DAW{}s by making implicit assumptions explicit. Once specified, \VC{} can be controlled automatically by the \DAW{} infrastructure, and violations can lead to meaningful error messages and graceful behaviour (e.g., termination or invocation of repair mechanisms). We provide a broad list of possible \VC{}s, classify them along multiple dimensions, and compare them to similar concepts one can find in related fields. We also provide a first sketch for \VC{}s' implementation into existing \DAW{} infrastructures.
\end{abstract}


\section{Introduction}
\label{sec:intro}

Data analysis workflows (\DAW{}s, or scientific workflows) are structured descriptions for scientific datasets' scientific analysis~\cite{dSFS+17,LAG+16}. \DAW{}s' usage becomes increasingly popular in all scientific domains as datasets grow in size, analyses grow in complexity, and demands grow in terms of speed of development, the throughput of analyses, reusability by others, and reproducibility of results~\cite{CBB+17,JSL+19,WvSL19}. A \DAW{} essentially is a program consisting of individual tasks (programs themselves) with their specific inputs and outputs and a specification of the dependencies between tasks. Executing a \DAW{} means scheduling its tasks on the available computational infrastructure in an order compatible with the data dependencies under some optimization constraints, such as minimal time-to-finish~\cite{Suk21,WWL19,YBR08}. When \DAW{}s are applied for the analysis of large datasets and are executed on clusters of distributed compute nodes, managing their execution also involves resource management, coordination of distributed computations, and file handling~\cite{isol22}. Such distributed executions typically rely on the availability of an infrastructure stack consisting of several components such as (distributed) file systems, resource managers, container managers, and runtime monitoring tools~\cite{Silva21}.

The interfaces and functionality of these components are not standardized and vary substantially between different systems~\cite{Silva21}. Therefore, \DAW{} developers often optimize their code to the used infrastructure (e.g., to the number and memory sizes of available compute nodes) and to the particular datasets they wish to analyze (e.g., by hard-coding the number of data partitions for concurrent execution). Furthermore, many tasks of typical \DAW{}s have been written by third parties, providing their specific functionality in a highly optimized manner, while others provide merely `glue' code for data transformation and filtering between original tasks~\cite{RLS+06, OGA+05}. As a result, real-life \DAW{}s are rather brittle artifacts. They are tightly bound to the infrastructure used during development, suffer from intricacies of the programs they embrace, and only work flawless for a narrow range of inputs. Changes in any of these aspects that violate hard-coded, undocumented design choices quickly lead to unforeseen situations such as: unnecessarily slow \DAW{} executions, underutilized resources, sudden runtime errors, straggler processes, meaningless log entries, non-terminating executions, overflows of buffers (memory, disk, log-space), etc. or, in the worst case, undetected faulty results~\cite{KKLS17}. The execution often stops with an arbitrary, undocumented, low-level error (`file not found', `core dumped', `timeout'); even meaningful error messages are often difficult to trace back to the broken task as execution happens on multiple nodes and logs are distributed and created at different levels, ranging from OS to resource managers, workflow engine, and task implementations. While some of these problems also occur in other software-related situations, they are aggravated in \DAW{}s due to their heavy reliance on external programs, generally very high resource requirements, long run times, and the complexity of coordinating distributed executions. Accordingly, reusing a \DAW{} on another infrastructure or for input data with differing properties often requires time-consuming adaptations~\cite{LFB+21,SBB+20}.

In this work, we propose validity constraints (\VC{}s) as a new primitive for \DAW{} languages that help to improve this situation. A \VC{} is a constraint that specifies a logical condition for a particular state or component of a \DAW{} execution. When a \VC{} evaluates to false (i.e., if the \VC{} is `broken'), the \DAW{} engine can issue a defined error message at a defined place. \VC{}s may, for instance, control properties of the input and intermediate data files (e.g., minimal or maximal file sizes), of the runtime environment (e.g., minimal available memory or threads), or of the individual task executions (e.g., maximal execution time). We propose to specify \VC{}s within the \DAW{} specification, i.e., as first-class objects of the \DAW{} program itself.

In the following, we first introduce a model for \DAW{}s (\secref{sec:definitions}) and then use this model to formally define general validity constraints (\secref{sec:formal_daws}). We present a broad list of different concrete types of \VC{}s (\secref{sec:vcs}) and classify these along multiple dimensions, namely the time points when they need checking, the objects they affect, the actions they may trigger, and the infrastructure component that should handle them (\secref{sec:vc-properties}). We relate \VC{}s to similar concepts in other fields, such as integrity constraints in databases or pre-/post-conditions in programming languages, and discuss previous works in the workflow community (\secref{sec:related-work}). Furthermore, we sketch a prototypical implementation of explicit \VC{}s in the state-of-the-art workflow engine Nextflow (\secref{sec:vc-impl}).

Throughout this work, we focus on simple \DAW{}s performing batch processing and leave an extension to data analysis over streams (e.g.,~\cite{FlowDB}) or to \DAW{}s including cycles or conditionals for future work.

\section{User Stories}
\label{sec:stories}

We collected a small set of typical problems users from different application domains ran into when using and porting \DAW{}s to another platform. Often, they stumbled over and had to solve validity constraints that were implicit and not explicit.

\subsection{Bioinformatics}
In bioinformatics research, we often modify or rewrite workflows, which requires developing short workflows performing RNAseq data treatment. We have to check the overall results for their quality and reasonability, but we are also interested in the effects our modifications may cause on different infrastructures performance-wise.

\paragraph*{Workflow Development---Empty and Faulty Files:}
During the development phase, errors can occur, and faulty files may be written. Such a situation does not necessarily interrupt the workflow directly because output files may exist.  Workflow engines typically do not assess a task's success by using the content of the output files. Identifying the wrongly behaving task in a distributed execution environment can become tedious and time-consuming for the user. For example, we once wanted to sort a file in the middle of a workflow but made a syntax error, which caused an empty output file. In this case, it would have been wonderful if we had a language with validity constraints that would help develop workflows and throw an error when an output file is empty, under a certain size, or does not contain specific characters. For such checks, the addition of monitoring tasks is necessary.

\paragraph*{Porting Workflows to New Infrastructures:}
We recently studied the impact of applying map-reduce on specific bioinformatics tools.
While porting a workflow on a heterogeneous distributed infrastructure, we observed a severely reduced workflow runtime. It turned out that only a few nodes could run tasks that should all run in parallel. We found the memory of most of the nodes too small to run these tasks. As a result, we changed the biological model to one with smaller input references and recomputed several experiments. Fortunately, that was possible in this case. But it may not be an option for biologists studying a specific specie, for example. If their reference genome file is too big for the memory of the nodes of a cluster, they would need to set up their workflow on another infrastructure. As such experiments can take a long time (up to 40 hours for treating only one sample), a way to know beforehand that the workflow cannot run (with the full degree of parallelism) on this infrastructure can save a lot of time and shared resources. Instead of executing a workflow by checking tasks' resource demands only late, when the execution reaches them, a basic overall resource check to stop the workflow from the beginning (before it arrives at the task that breaks the workflow) would be preferable.

\begin{quote}\emph{``It would be wonderful if I had a language with validity constraints that would help me develop workflows or port them to a new infrastructure. I would have needed constraints that throw an error when an output file is empty, is under a certain size or does not contain specific characters such as those contained in a specific header. Additionally, some constraints that would stop the workflow from the beginning, before it arrives at the task that requires too much resources, would be very helpful.''}\hfill-- Ninon De Mecquenem
\end{quote}

\subsection{Materials Science}

The novel materials database laboratory (\NOMAD{}) is a database that hosts hundreds of millions of material science simulation results, specifically density functional theory (\DFT{}) simulations~\cite{DS+19}. These results can be expensive to generate, sometimes taking several hundred \GPU{} hours to compute~\cite{ED+11}. As such, the community realized it is vital to share the results to avoid recomputation and to allow for the creation of large datasets for machine learning and data analysis applications~\cite{SAB22, AAB21}.

Users can upload data (input/output files) from different density functional theory simulation programs (e.g., \VASP{}, \exciting\, \FHIaims{}) via the terminal or their browser~\cite{GKM14,BRK22,H08}. The upload process data workflow has a clear need for validity constraints. Each simulation program has an associated parser to parse the simulation input/output files that the user uploads~\cite{VDS19}. A common pain point is that these simulation programs get updated frequently, and the output files' format changes or new data fields are added. The \NOMAD{} developers implemented \VC{}s in the upload \DAW{} to check for the existence of specific file names, extension types, and some properties of the input/output files our parser expects~\cite{GKM14}.  For the \DFT{} code \exciting\, for example, we expect files named \texttt{INFO.out} and \texttt{input.xml} files and particular key-value pairs, such as the `total energy' key and its associated floating point value. This means our \DAW{} first runs a validity constraint at the setup of the upload process to check that required files exist. If these simulation files do not exist the \DAW{} returns an error message to the user that the upload failed since no files that could be parsed were found. The upload process also triggers a resource availability validity constraint that checks whether the user has used up the amount of storage space allocated to every user of \NOMAD{}.

\begin{figure}[h]
\centering
\includegraphics[width=.75\linewidth]{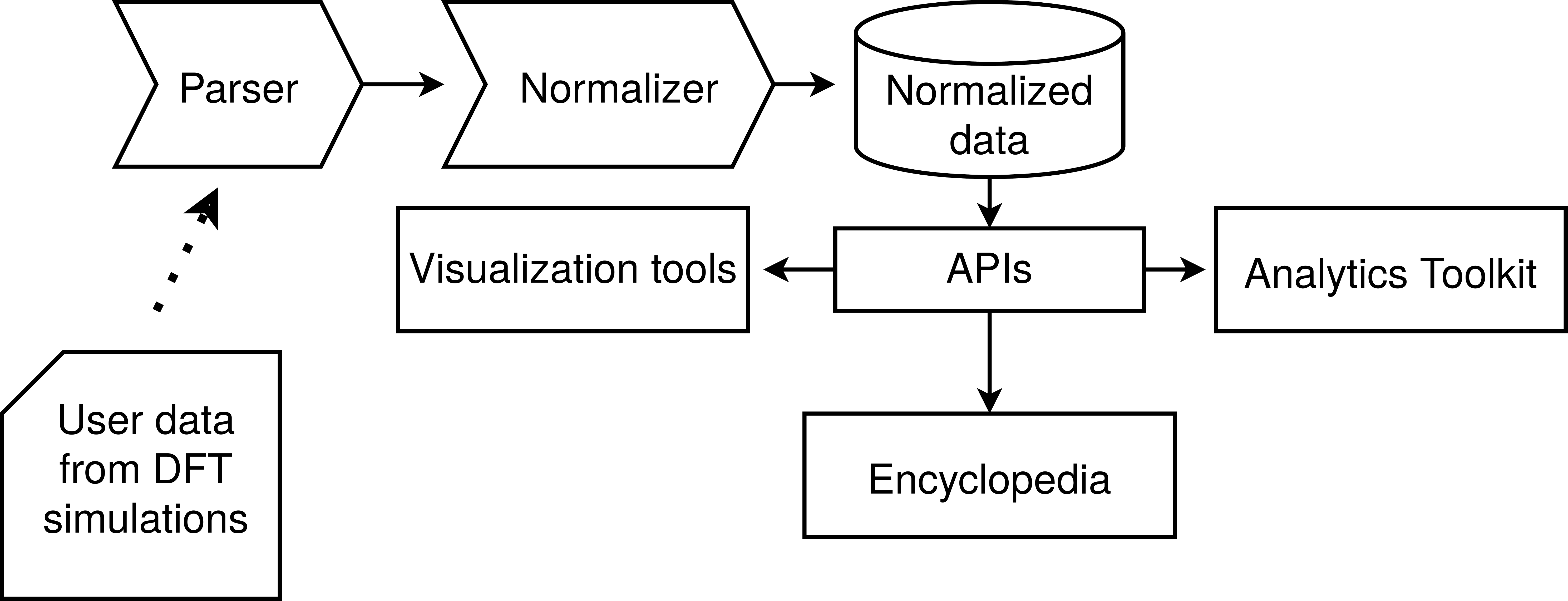}
\caption{Overview of the \NOMAD{} upload workflow. Users upload data from a specific \DFT{} simulation code that is then parsed and normalized to make results from different \DFT{} codes comparable.}
\label{fig:nomad_overview}
\end{figure}

If however, the uploaded files do contain simulation files that can be parsed the \DAW{} then runs a resource check validity constraint. The parsing process can take considerable resources depending on the simulation settings and the \DAW{} only executes the parsing once Kubernetes can allocate sufficient computing resources on the server. After parsing raw values from the uploaded files, a routine called the `normalizer' is applied, which converts parsed values and simulation settings to standard units and standardized terms. For instance, two simulation programs might use different words for the same parameter and \NOMAD{} needs to standardize this (e.g., two different names for the same \DFT{} functional). The normalizer also implements validity constraints on these parsed values to ensure they are reasonable. For instance, the normalizer checks that categorical property to be from a list of expected values or that a floating point value is within a reasonable range (e.g., the band gap of the material must be non-negative). The \DAW{} for \NOMAD{} can be seen in \figref{fig:nomad_overview}.

When we have too many atoms in a unit cell of an input geometry, we observe another weakness of the \DAW{} regarding the crystal structure classification, which is performed in the normalizer step. Large input files with many atoms in the unit cell are common in studies that investigate the effect of impurities on the electronic structure of crystalline materials~\cite{CND20}. Such a situation causes the crystal structure classifier to take a very large amount of computational resources. Currently, we use a timeout validity constraint that stops the classification if the classification takes too long. What might help, however, is to implement a validity constraint that, based on the number of atoms in the unit cell, decides whether to skip the crystal structure classification for unit cells where the number of atoms exceeds a certain threshold. This could help avoid wasting resources on trying to classify systems that are very likely to trigger the timeout validity constraint during classification. Such a threshold could be determined using a logistic regression model trained on previous uploads and workflow executions. Alternatively, the threshold could be dependent on the resources available for computation. In this case, we envisage \VC{}s that are chained together. Meaning, first we check if resources are somewhat limited and if so, we call a \VC{} that checks if the number of atoms in the unit cell is too large. If this is true, it avoids the crystal structure classification all together.

A further valuable addition to the \NOMAD{} upload \DAW{} would be a check for a metamorphic relation between input settings of the simulation and output results of the simulation to predict data quality~\cite{VDS19}. Simulations uploaded to \NOMAD{} usually come from different applications. They could be, for instance, super high precision ground state calculations (e.g., using very large basis set sizes) or simulations to find heat and transport properties using cheap calculations (small basis set size), resulting in data of varying precision~\cite{CTB22}. Current research has been on using machine learning models to add prediction intervals that act as error bars to the data results parsed by \NOMAD{}~\cite{SCG23}. Such annotations would help users better understand how precise and, therefore, how qualified results from \NOMAD{} are for a particular use-case. For example, a formation energy calculated with high precision settings would have small prediction intervals. It shows the \NOMAD{} end user (e.g., an experimentalist) that this result does not need a recalculation.

There are currently many \VC{}s already integrated into the \NOMAD{} \DAW{} and some that we would like to add as discussed above. What would be useful is to have a language that can clearly present the \VC{}s of the \DAW{} in a graphical form in a visual format. We believe this would help end-users and developers better understand the \DAW{} especially when the process fails due to one of these constraints.

\subsection{Earth Observation}

In the field of Earth Observation (\EO{}), we typically work with large volumes of satellite images, often over large areas (from federal states to continents, sometimes even global) and across long time series (up to 40 years). Our workflows often consist of a preprocessing step to convert the `raw' data into a more analysis-ready form. This preparation typically includes the detection of clouds and their shadows, eliminating atmospheric and other radiometric distortions, and often rearranging the data into so-called data cubes for improved data management and efficiency. This processing step usually runs on each individual satellite image separately, allowing image-based parallelism. Subsequently, workflows generally use map/reduce operations on spatial partitions of the data cube, or even sub-partitions that need specific parts of the images. Typically, all available data for some time period is reduced (averaged in the simplest case) to generate gap-free predictive features. The feature vectors at the locations of reference data are extracted, a machine learning model trained with some response variable (e.g., land cover labels or tree height values), and the model applied to the feature images to generate a wall-to-wall prediction of the respective response variable (i.e., generating a map). A validation procedure typically follows. 

Different \IT{} resources, such as \CPU{}s, \RAM{}, and \IO{} bandwidth, typically constrain various components of this generic workflow. Depending on the particular workflow, the analyzed data, user parameterization, as well as characteristics of the hardware, the limiting factor might be different each time. For example, a workflow that efficiently runs with Landsat data might fail when switching to Sentinel-2 data (more \RAM{} needed) even when executed with the same parameterization on the same system. Another example would be a workflow that efficiently reads data but would quickly become input-limited when switching from an \SSD{}- to an \HDD{}-based platform, or another \RAID{} configuration. In a noteworthy instance, we encountered an extreme worst-case scenario in which processes would sporadically become defunct while unzipping files from tape storage. As a result, our job would come to a complete halt after a certain period. Resolving this issue required a specific modification of our workflow, including the addition of a hard-coded timeout. Moreover, we had to first copy data from tape to warm storage before executing the workflow again. 

Consequently, a one-fits-all default parameterization is usually not feasible, and many user parameters may exist that can tweak the behavior of the workflow. For example, the \FORCE{} software~\cite{frantz2019} includes parameters to fine-tune partition sizes, reduce the number of parallel processes, or to increase the multithreading to multiprocessing ratio when \RAM{} becomes an issue. However, achieving optimal parameterization needs a deep understanding of the workflow, the underlying data, and their effects on system resources. Additionally, a solid understanding of the platform is essential for effective parameterization. Therefore, the presence of validity constraints capable of identifying common patterns of excessive resource usage, such as idle \CPU{}s or high network latency in \IO{}-limited scenarios, or memory swapping leading to generic `killed' messages in \RAM{}-limited situations, would significantly aid in transferring workflows from one system to another. Additionally, it would facilitate a smoother onboarding process for new users, reducing the learning curve required.

\section{Fundamentals}
\label{sec:definitions}

In this section, we formally define a \DAW{} and its execution steps, sketch an abstract infrastructure for executing \DAW{}s in a  distributed system, and introduce scheduling as the process of executing a \DAW{} on an infrastructure. Based on these models, we then define two types of general validity constraints (static and dynamic) as new first-class primitives for \DAW{} specification languages, and use them to derive the concept of valid and correct \DAW{} execution.

Our \DAW{} semantic is simple by intention; its purpose is to lay the grounds for the following sections, which will precisely define the connection between elements of a \DAW{} and \VC{}s and the impact that \VC{}s may have on \DAW{} execution. Conceptually, our semantics is similar to Petri-Nets~\cite{diaz09} and dataflow languages~\cite{JHM04}. Elaborated semantics of real workflow systems have been described elsewhere (e.g.,~\cite{SHMG10,ZBM+09}); \cite{LSV98} gives a nice overview of different formal models in distributed computation.

\subsection{A Formal Model of \DAW{}s}

We define a logical \DAW{} (see below for the distinction to physical \DAW{}s) as follows.

\begin{definition}[\emph{Logical \DAW{}}]
A logical \DAW{} $W$ is a directed acyclic graph
\begin{equation}
W=(T, D, L, \phi, t_s, t_e)
\label{eq:logical-daw}
\end{equation}
\end{definition}
\noindent
where $T$ is the set of tasks,  $D=\{(t',t) \in T^2 \}$ is the set of dependencies between pairs of tasks, $L$ is a set of labels,  $\phi: D \to L$ is a function assigning labels to dependencies, $t_s \in T: \nexists (t',t_s) \in D$ is the start task, and $t_e \in T: \nexists (t_e, t') \in D$ is the end task. Intuitively, tasks are the programs to be executed for performing individual analysis steps, while dependencies model the data flow between tasks. The dependencies' label is an abstract representation of the specific data that is exchanged between two tasks. The start task $t_s$ does not depend on any other task and initiates the first steps of the analysis by sending the \DAW{}'s input data to its dependent tasks. Similarly, the end task $t_e$ has no dependent tasks, and the labels of its incoming dependencies represent the results of the \DAW{}. Figure \ref{fig:1} (upper part) shows a graphical representation of an example \DAW{} consisting of six tasks plus start and end tasks; arcs represent dependencies.

\DAW{}s are executed by running their tasks in an order in which at all times all dependencies are satisfied. To formally define this semantics, we introduce the notation of the state of a \DAW{} and, later, that of valid states.

\newcommand{\sF}{\textsf{F}}
\newcommand{\sR}{\textsf{R}}
\newcommand{\sO}{\textsf{O}}

\begin{definition}[\emph{State of a \DAW{}}]
The state $S^{W}$ of a \DAW{} $W$ is a function that assigns each task in the set $T$ to one of three possible states:
\begin{equation}
S^{W}: T \to \{\sF{},\sR{},\sO{}\}
\label{eq:state-of-daw}
\end{equation}
\end{definition}\noindent
Here, \sF{} means `finished', \sR{} means `ready' and \sO{} means `open'.

\begin{definition}[\emph{Valid States}]
The state $S^{W}$ of a \DAW{} $W$ is \emph{valid}, iff the following conditions hold:

\begin{itemize}
    \item $S(t_s)=\sF{}$,
    \item $\forall (t', t) \in D$
    $\forall t \in T$ with $(t',t) \in D$: if $S(t)=\sR{}$, then $S(t')=\sF{}$,
    \item $\forall (t', t) \in D$
    $\forall t \in T$: If $\forall t'$ with $(t',t) \in D$: $S(t')=\sF{}$, then $S(t)=\sR{}$, and
    \item for all other $t \in T: S(t)=\sO{}$.
\end{itemize}
The initial state $S_0$ of a \DAW{} $W$ is
the state in which (1) the start task is finished: $S_0(t_s)=\sF$, (2) all tasks $t'$ depending on $t_s$ are ready: $S_0(t')=\sR{}$, and (3) all other tasks have state `open'.
\end{definition}

Intuitively, these rules guarantee that: (a) the start task is always in the `finished' state; (b) a task is `ready' only when all its predecessors are `finished'; (c) any task with all its predecessors `finished' has state `ready'; and (d) all tasks not fulfilling any of the previous three conditions (a)--(c) are in the `open' state. The initial state of a \DAW{} is the state before its execution.

\subsection{\DAW{} Infrastructure and Execution Semantics}
\label{sec:daw-infrastructure}

Based on a \DAW{}'s state, we next define the semantics of a \DAW{} execution.

\begin{definition}[\emph{Execution of a \DAW{}}]
An execution $E$ of \DAW{} $W$ is a sequence of states $E =
\langle S_0,\ldots,S_n\rangle$ such that (a) $S_0$ is the \DAW{}'s initial state, (b) all $S_i, i\in\{0,\ldots,n\}$, are valid, and (c) for all steps $S_i, S_j$ with $j=i+1$, it holds that
\begin{itemize}
    \item If $S_i(t)=\sF{}$, then $S_j(t)=\sF{}$
    \item If $S_i(t)=\sR{}$, then $S_j(t) \in \{\sF{},\sR{}\}$
    \item If $S_i(t)=\sO{}$, then $S_j(t) \in \{\sO{},\sR{}\}$
    \item There exists at least one $t\in T$ where $S_i(t) \neq S_j(t)$.
\end{itemize}
We say that an execution $E$ of a \DAW{} $W$ has executed $W$ when $S_n(t_e)=\sF{}$.

\end{definition}

\newcommand{\once}{$\overset{\,_1~}{\to}$}
\newcommand{\cont}{$\leadsto$}
\newcommand{\period}{$\circlearrowright$}
\newcommand{\event}{$\mapsto$}
\newcommand{\before}{$\to$\kern-.18em$[$\,}
\newcommand{\after}{\,$]$\kern-.18em$\leftarrow$}
\newcommand{\during}{$[$\kern-.12em$\leadsto$\kern-.18em$]$}
\newcommand{\discr}{$\downarrow$}
\newcommand{\trigger}{$\Rsh$}
\newcommand{\yes}{$\checkmark$}
\newcommand{\no}{$\times$}
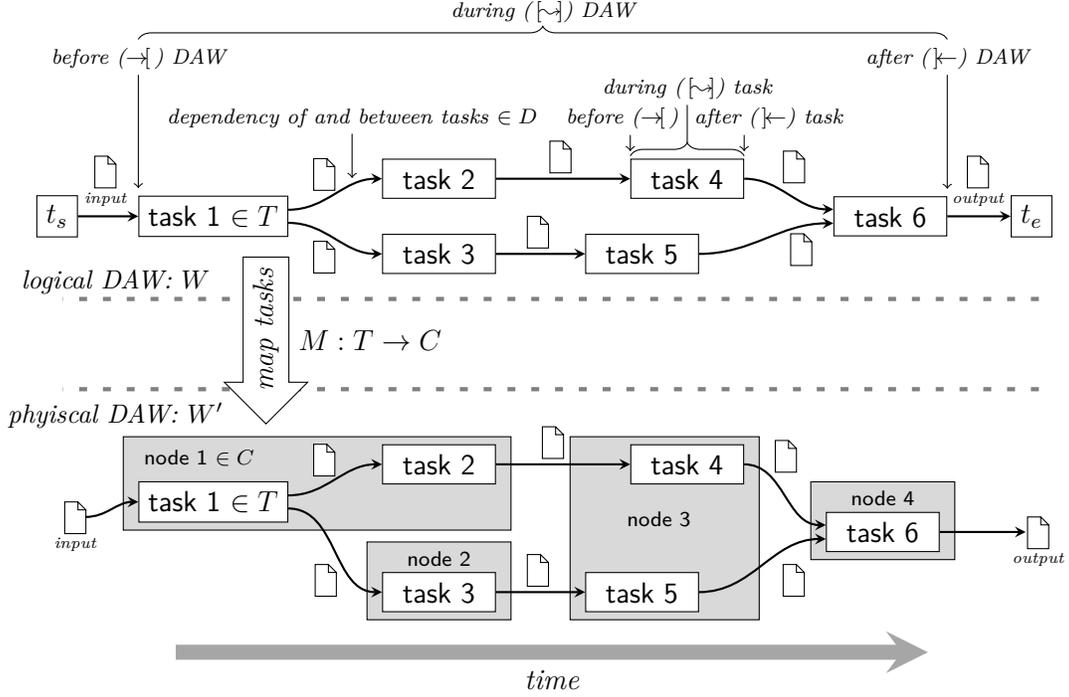
\begin{figure}[t!]
\centering\begin{tikzpicture}[
  task/.style={rectangle, draw=black, fill=white, minimum width=15mm,
    align=center, font=\sffamily},
  label/.style={inner sep=2pt, font=\itshape},
  >=stealth,
  storagenode/.style={cylinder, shape border rotate=90, shape aspect=.2, draw=black, fill=gray!30!white, align=center, font=\sffamily}
]

\newcommand{\filesymbol}[2]{\filldraw[white,draw=black,text=black] #1
    -- ++(-5pt,0) 
    -- ++(0,-12pt) 
    -- node[below=-2pt,font=\tiny\itshape] {#2} ++(8pt,0) 
    -- ++(0,9pt) 
    -- cycle 
    -- ++(0,-3pt) -- ++(3pt,0); 
}

\coordinate (input) at (-2,-.3);
\draw (input) +(.2,-.2) node[task,anchor=east,minimum width=1mm] (ts) {$t_s$};
\node[task] at (0,-.5) (t1) {task 1 $\in T$};
\node[task] at (3,0) (t2) {task 2};
\node[task] at (3,-1) (t3) {task 3};
\node[task] at (6.3,0) (t4) {task 4};
\node[task] at (5.7,-1) (t5) {task 5};
\node[task] at (9,-.5) (t6) {task 6};
\coordinate (output) at (11,-.3);
\draw (output) +(-.4,-.2) node[task,anchor=west,minimum width=1mm] (te) {$t_e$};

\filesymbol{([xshift=6mm,yshift=6mm]input)}{input}
\filesymbol{([xshift=-8mm,yshift=6mm]output)}{output}

\draw[->,thick] (ts) to[out=0,in=180] (t1);
\draw[->,thick] (t1.5) to[out=0,in=180] coordinate[pos=.4] (t1tot2) coordinate[pos=.7] (dependency) (t2);
\draw[->,thick] (t1.355) to[out=0,in=180] coordinate[pos=.4] (t1tot3) (t3);
\draw[->,thick] (t2) to coordinate[above=5mm] (t2tot4) (t4);
\draw[->,thick] (t3) to coordinate[above=5mm] (t3tot5) (t5);
\draw[->,thick] (t4) to[out=0,in=180] coordinate[pos=.6] (t4tot6) (t6.173);
\draw[->,thick] (t5) to[out=0,in=180] coordinate[pos=.8] (t5tot6) (t6.187);
\draw[->,thick] (t6) to[out=0,in=180] (te);

\filesymbol{(t1tot2 |- t2.north)}{}
\filesymbol{([yshift=-1mm]t1tot3 |- t3.north)}{}
\filesymbol{(t2tot4)}{}
\filesymbol{(t3tot5)}{}
\filesymbol{([yshift=1mm]t4tot6 |- t4.north)}{}
\filesymbol{(t5tot6 |- t5.north)}{}

\node[label] at (dependency |- 0,.8) (dependencylabel) {\scriptsize dependency of and between tasks $\in D$};
\draw[->,>=to] (dependencylabel) -> ([yshift=1mm]dependency);

\node[label] at (t1.west |- 0,1.6) (beforedaw) {\scriptsize before (\before) DAW};
\draw[->,>=to] (beforedaw) -> ([yshift=1mm]t1.north west);
\node[label] at (t6.east |- 0,1.6) (afterdaw) {\scriptsize after (\after) DAW};
\draw[->,>=to] (afterdaw) -> ([yshift=1mm]t6.north east);
\draw[decorate,decoration={brace,amplitude=5pt,raise=1pt}] (beforedaw.north) -- (afterdaw.north) node[label,pos=.5,yshift=4mm] (indaw) {\scriptsize during (\,\during) DAW};

\node[label] at (t4.west |- 0,.8) (beforetask) {\scriptsize before (\before)~~~};
\draw[->,>=to] (beforetask) -> ([yshift=1mm]t4.north west);
\node[label] at (t4.east |- 0,.8) (aftertask) {\scriptsize~~~~~~after (\after) task};
\draw[->,>=to] (aftertask) -> ([yshift=1mm]t4.north east);

\draw[decorate,decoration={brace,amplitude=5pt,raise=1pt,mirror}] (t4.north east) -- (t4.north west);
\node[label] at (t4.north |- 0,1.2) (intask) {\scriptsize during (\,\during) task};
\draw (intask) -> ([yshift=2mm]t4.north);

\draw[->,line width=2mm,gray!70!white,text=black] (-.5,-6.3) -- node[label,below=.8mm] {time} (9.5,-6.3);

\draw[gray, ultra thick, loosely dashed, dash phase=1mm] (input |- 0,-1.6) coordinate (separateleft) -- (output |- 0,-1.6);
\node[label,anchor=south ] at ([xshift=7mm]separateleft) {\small logical DAW: $W$};

\draw[gray, ultra thick, loosely dashed, dash phase=1mm] (input |- 0,-2.8) coordinate (separate2left) -- (output |- 0,-2.8);
\node[label,anchor=north ] at ([xshift=7mm,yshift=-1mm]separate2left) {\small
  phyiscal DAW: $W'$};

\node[draw,fill=white,single arrow,rotate=90,shape border rotate=180] at ([xshift=7mm,yshift=-4mm]separateleft -| t1) (mapping) {\emph{map~tasks}};
\node[anchor=west,xshift=3mm,yshift=-1.5mm] at (mapping) {$M: T \to C$};


\coordinate (pinput1) at (-2.2,-4.3);
\node[task] at (0,-4.3) (pt1) {task 1 $\in T$};
\node[task] at (3,-3.8) (pt2) {task 2};
\node[task] at (3,-5.5) (pt3) {task 3};
\node[task] at (6.3,-3.8) (pt4) {task 4};
\node[task] at (5.7,-5.5) (pt5) {task 5};
\node[task] at (8.9,-4.7) (pt6) {task 6};
\coordinate (poutput1) at (11,-4.5);
\coordinate (poutput2) at (11,-5.9);

\filesymbol{([xshift=4mm]pinput1)}{input}

\filesymbol{(poutput1)}{output}

\draw[->,thick] ([xshift=5mm,yshift=-2mm]pinput1) to[out=0,in=180] (pt1);

\draw[->,thick] (pt1.5) to[out=0,in=180] coordinate[pos=.4] (t1tot2) (pt2);
\draw[->,thick] (pt1.355) to[out=0,in=180] coordinate[pos=.4] (t1tot3) (pt3);

\draw[->,thick] (pt2) to[out=0,in=180] coordinate[above=5mm] (t2tot4) (pt4);
\draw[->,thick] (pt3) to[out=0,in=180] coordinate[above=5mm] (t3tot5) (pt5);

\draw[->,thick] (pt4) to[out=0,in=180] coordinate[pos=.6] (t4tot6) (pt6.173);
\draw[->,thick] (pt5) to[out=0,in=180] coordinate[pos=.6] (t4tot6) (pt6.187);
\draw[->,thick] (pt6) to[out=0,in=180] ([xshift=-2mm] poutput1 |- pt6);

\filesymbol{(t1tot2 |- pt2.north)}{}
\filesymbol{([yshift=1mm]t1tot3 |- pt3.north)}{}
\filesymbol{([xshift=-1mm]t2tot4)}{}
\filesymbol{(t3tot5)}{}
\filesymbol{([xshift=2mm,yshift=.5mm]t4tot6 |- pt4.north)}{}
\filesymbol{([xshift=-1mm,yshift=1mm]t5tot6 |- pt5.north)}{}


\begin{scope}[on background layer]
\filldraw[gray!30!white,draw=black] ([xshift=-2mm,yshift=-1mm]pt1.south west) rectangle ([xshift=2mm,yshift=1mm]pt2.north east);
\node[label,font=\sffamily,anchor=north west] at (pt1.south west |- pt2.north east) {\scriptsize node 1 $\in C$};

\filldraw[gray!30!white,draw=black] ([xshift=-2mm,yshift=-1mm]pt3.south west) rectangle ([xshift=2mm,yshift=4mm]pt3.north east);
\node[label,font=\sffamily] at ([yshift=4.5mm]pt3 |- pt3) {\scriptsize node 2};

\filldraw[gray!30!white,draw=black] ([xshift=2mm,yshift=1mm]pt4.north east) rectangle ([xshift=-2mm,yshift=-1mm]pt5.south west |- pt5.south);
\node[label,anchor=south,font=\sffamily] at (pt5.50 |- pt6.east) {\scriptsize node 3};

\filldraw[gray!30!white,draw=black] ([xshift=-2mm,yshift=-1mm]pt6.south west) rectangle ([xshift=2mm,yshift=4mm]pt6.north east);
\node[label,font=\sffamily] at ([yshift=4.5mm]pt6 |- pt6) {\scriptsize node 4};
\end{scope}
\end{tikzpicture}%
\caption{A logical (upper part) \DAW{} and its physical counterpart (lower part).}
\label{fig:1}
\end{figure}

Intuitively, the execution of a \DAW{} progresses by iteratively executing tasks that are ready to run. After execution, their state switches to \sF{}; tasks in state \sO{} must first proceed to state \sR{} before they can run. We make no assumptions regarding the order in which tasks that are ready to run at the same state are executed, nor do we assume only one task executes per execution step. But we do require at least one task to change its state between successive \DAW{} states. Note that this change may be purely logical, by switching some task's state from \sO{} to \sR{}.

\begin{figure}[t!]
\centering\begin{tikzpicture}[
  task/.style={rectangle, draw=black, fill=white, minimum width=15mm,
    align=center, font=\sffamily},
  label/.style={inner sep=2pt,font=\itshape,text height=1.5ex, text depth=.25ex},
  >=stealth,
  storagenode/.style={cylinder, shape border rotate=90, shape aspect=.15, draw=black, fill=white, align=center, font=\sffamily}
]

\coordinate (input) at (-2,-.3);
\node[label] at (0,0.1) (daw-sl) {DAW specification language};
\node[label] at (0,-1) (daw-ee) {DAW execution environment (EE)};
\node[label,anchor=west,xshift=-7mm] at (daw-ee.west |- 0,-2) (scheduler) {scheduler (S)};
\node[label,anchor=east,xshift=7mm] at (daw-ee.east |- 0,-2) (rm) {resource manager (RM, M)};

\draw[->,thick] (daw-sl) to node[right,label] {DAW compiler (comp.)} (daw-ee);
\draw[->,thick] (daw-ee) to[out=230,in=50] (scheduler);
\draw[->,thick] (daw-ee) to[out=290,in=120] (rm);
\draw[->,thick] (scheduler) to (rm);

\filldraw[gray!30!white,draw=black] (scheduler.west |- 0,-3) rectangle
coordinate (node1) +(1.6,-1.9);
\filldraw[gray!30!white,draw=black] (rm.east |- 0,-3) rectangle
coordinate (node3) +(-1.6,-1.9);

\path (scheduler.west) -- coordinate[xshift=-8mm] (middle) (rm.east);
\filldraw[gray!30!white,draw=black] (middle |- 0,-3) rectangle
coordinate (node2) +(1.6,-1.9);

\coordinate (node1top) at (node1 |- 0,-3);
\coordinate (node2top) at (node2 |- 0,-3);
\coordinate (node3top) at (node3 |- 0,-3);

\node[label,font=\sffamily] at ([yshift=7mm]node1) {\scriptsize node 1};
\node[label,font=\sffamily] at ([yshift=7mm]node2) {\scriptsize node 2};
\node[label,font=\sffamily] at ([yshift=7mm]node3) {\scriptsize node 3};

\draw[->,thick] (rm) to[out=210,in=40] (node1top);
\draw[->,thick] (rm.240) to[out=250,in=60] (node2top);
\draw[->,thick] (rm) to[out=290,in=100] (node3top);

\node[storagenode] at ([yshift=-4mm]node1) (storage1) {\scriptsize storage 1};
\node[storagenode] at ([yshift=-4mm]node2) {\scriptsize storage 2};
\node[storagenode] at ([yshift=-4mm]node3) (storage3) {\scriptsize storage 3};

\draw ([xshift=-1mm,yshift=1mm]storage1.north -| scheduler.west) rectangle
      ([xshift=6mm,yshift=-1mm]storage3.south -| rm.east);

\node[label,anchor=north,xshift=4mm] at (storage3.east) (DFS) {\rotatebox{90}{DFS}};

\draw[->,thick] (rm.-2) to[out=340,in=40] ([xshift=6mm]rm.east |- storage3.10);
\end{tikzpicture}%
\caption{A simple \DAW{} infrastructure architecture.}
\label{fig:daw-infrastructure-architecture}
\end{figure}
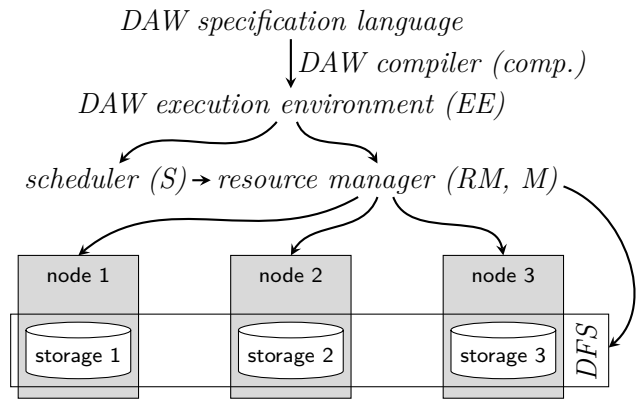

Logical \DAW{}s are abstract objects. However, in real life, a \DAW{} execution requires the start of programs representing a workflow task on a particular node of the available cluster and the management of the inputs and outputs of these programs. \figref{fig:daw-infrastructure-architecture} depicts a light architecture of the components involved in such a \DAW{} execution. It encompasses  the \DAW{} specification in a proper \DAW{} language and its compiler (\emph{comp.}), the \DAW{} engine steering the \DAW{} execution (\EE{}), a scheduler performing the task-to-node assignments (\Ss{}), a resource manager and monitoring system controlling the resource assignment and task execution at the global and local level (\RM{}, \M{}), the individual nodes for executing tasks, and a distributed file system for data exchange between tasks (\DFS{})\footnote{Of course, other means of data exchange, such as in-memory channels, or mounting of remote file systems are possible as well.}. Clearly, many other architectures are possible, but for the sake of this work, such an idealized architecture suffices and allows later determining the responsible component to control a particular type of \VC{}. With such an architecture in mind, we can now define a physical \DAW{}.

\begin{definition}[\emph{Physical \DAW{}}]
Given a logical \DAW{} $W$ and a set $C$ of compute nodes interconnected by a network, the physical \DAW{} $W' = (W, M)$ augments $W$ with a function $M: T \to C$ that maps every task to a compute node.
\end{definition}

$C$ is an abstract representation of a compute cluster, whereas $M$ is an assignment (schedule) that maps tasks to nodes. The definition of a \emph{\DAW{} execution} $E$ can be naturally extended from logical \DAW{}s to physical \DAW{}s. Here, assignment $M$ determines where a task whose state switches from \sR{} to \sF{} is executed.

Figure \ref{fig:1} shows an example of the transition of a logical \DAW{} to a physical \DAW{}. In the physical \DAW{}, each logical task is assigned to a node for execution, and each dependency is implemented as communication between nodes.

\section{A Formal Definition of \VC{}s for \DAW{}s}
\label{sec:formal_daws}

Having introduced logical and physical \DAW{}s and their execution semantics, we can now define validity constraints (\VC{}s) as logical formulas over the components of a \DAW{} infrastructure and of a \DAW{} execution, i.e., tasks, dependencies, executions, and schedules. Different \VC{}s will address various properties of these components (see \secref{sec:vcs}).

\begin{definition}[\emph{Properties}]
Let $W=(T, D, L, \phi, t_s, t_e)$ be a \DAW{} and $C$ be a cluster, i.e., a set of compute nodes. We model arbitrary properties of elements of $T$, $D$, and $C$ as property functions:

\begin{itemize}
    \item $P_T$ is a function that assigns properties to tasks from $T$,
    \item $P_D$ is a function that assigns properties to labels of dependencies from $D$, and
    \item $P_C$ is a function that assigns properties to nodes from $C$.
\end{itemize}
\end{definition}

We make no assumptions on the specific nature of such properties, such as data type or number of parameters they take.  In \secref{sec:vcs}, we will give a diverse list of concrete (static or dynamic) properties that we consider particularly useful for \DAW{} management.

\subsection{Validity Constraints}
\label{sec:vc-def}

We discern two types of validity constraints: Static \VC{}s address static properties, i.e., properties which always have the same value for a given property of a task/node/label, or while dynamic \VC{}s address dynamic properties, i.e., properties whose value may change during a \DAW{} execution.

\begin{definition}[\emph{Static \VC{}}]
Let $W=(T, D, L, \phi, t_s, t_e)$ be a \DAW{} and $C$ be a cluster. Let $P_T$, $P_D$, and $P_C$ be their respective property functions. A static validity constraint $v$ is a formula of any of the following forms (with $\mathcal{C}$ being an arbitrary constant and $\boxdot \in \{=, <, >, \leq, \geq\}$):

\begin{itemize}
    \item \makebox[26mm][l]{$P_T(t) \; \boxdot \; \mathcal{C}$} for any task $t$ from $T$,
    \item \makebox[26mm][l]{$P_D(\phi(d)) \; \boxdot \; \mathcal{C}$} for the label $\phi(d)$ of any dependency $d$ from $D$, and
    \item \makebox[26mm][l]{$P_C(c) \; \boxdot \; \mathcal{C}$} for any node $c$ from $C$,
\end{itemize}
\end{definition}

We call \VC{}s of these three forms \emph{static} because they are independent of a \DAW{}'s execution. Intuitively, this implies that they must evaluate to the same value all the time before, during, and after a workflow's execution. An example of a static \VC{} would be the minimum size of main memory that must be available on a node on which a given task is about to be scheduled, or the availability of at least one node with a given minimum memory size within the cluster.

The second class of \VC{}s are \emph{dynamic} \VC{}s, which constrain properties of tasks, dependencies, or nodes that may change during a \DAW{}'s execution. For instance, executing a particular task in the middle of a \DAW{} may require the existence (or minimal size of a certain format) of a file that is created by previous steps in the very same \DAW{} execution. Introducing such dynamic \VC{}s requires first defining the scope of a step within an execution.

\begin{definition}[\emph{Scope of execution steps}]
Let $W=(T, D, L, \phi, t_s, t_e)$ be a \DAW{}, $E$ an execution of $W$, and $M$ a schedule for $W$ over a cluster $C$. Let $P_T$, $P_D$, and $P_C$ be their respective property functions. Furthermore, for a step $s$ from $E$, let $X(s) \subset T$ be the set of tasks of $W$ that are executed in this step, i.e., whose state changes from \sR{} to \sF{}; let $Y^i(s) \subset D$ be the set of dependencies from which tasks in $X(s)$ depend (incoming edges of all tasks executed in this step, each representing an input for a task); let $Y^o(s) \subset D$ be the set of dependencies outgoing from tasks in $X(s)$ (outgoing edges, each representing an output of a task); and let $C(s,t) \in C$ be the node on which $M$ schedules task $t \in X(s)$ for execution. We call the tuple $scope(s) = (X(s), Y^i(s), Y^o(s), C(s,t))$ the scope of step $s$.
\end{definition}

\begin{definition}[\emph{Dynamic \VC{}}]
A dynamic validity constraint $V$ over the scope $scope(s) = (X(s), Y^i(s), Y^o(s), C(s,t))$ of a step $s$ of a valid execution of a \DAW{} $W$ is a formula of any of the following forms (with $\mathcal{C}$ being an arbitrary constant and $\boxdot \in \{=, <, >, \leq, \geq\}$):

\begin{itemize}
    \item \makebox[26mm][l]{$P_T(s) \; \boxdot \; \mathcal{C}$} for a property of any task $t \in X(s)$,
    \item \makebox[26mm][l]{$P_C(s,t) \; \boxdot \; \mathcal{C}$} for a property of node $c=C(s,t)$,
    \item \makebox[26mm][l]{$P_D^i(s) \; \boxdot \; \mathcal{C}$} for a property of a label $\phi(d)$ with $d \in Y^i(s)$, and
    \item \makebox[26mm][l]{$P_D^o(s) \; \boxdot \; \mathcal{C}$} for a property of a label $\phi(d)$ with $d \in Y^o(s)$.
\end{itemize}

\end{definition}

\subsection{Correct \DAW{}s and Correct \DAW{} Executions}

We defined \VC{}s as logical rules over the components (static) or steps (dynamic) of a \DAW{} that evaluate to either true or false. However, we so far did not describe what consequences the evaluation of a such a rule should have. Intuitively, \VC{}s are intended as sensors, where an evaluation to `true' implies that no issue was detected, while an evaluation to `false' points to a concrete problem. To formally define this intuition, we next introduce the notion of correct \DAW{} setups, where a \DAW{} setup is a combination of a concrete \DAW{} and a concrete cluster on which it should be executed, and correct \DAW{} executions. Recall that we discerned static \VC{}s, whose evaluation always returns the same result for a given combination of \DAW{} and cluster, from dynamic \VC{}s, which are defined over \emph{executions} of \DAW{}s.

\begin{definition}[\emph{Correct \DAW{} setup}]
Let $W=(T, D, L, \phi, t_s, t_e)$ be a \DAW{}, $C$ a cluster, i.e., a set of interconnected compute nodes, and $V$ a set of static \VC{}s over $P_T$, $P_D$, and $P_C$. We say that the tuple $(W, C)$ is correct with respect to $V$ when all constraints in $V$ evaluate to true.
\end{definition}

We will omit $C$ when it is clear from the context and simply say that $W$ is correct for $V$.

\begin{definition}[\emph{Correct \DAW{} execution}]
Let $W=(T, D, L, \phi, t_s, t_e)$ be a \DAW{}, $E$ an execution of $W$, $M$ a schedule for $W$ over a cluster $C$, $V^s$ be a set of static \VC{}s over $W$ and $C$, and $V^d$ be a set of dynamic \VC{}s over $W$, $E$ and $C$. We say that $E$ is correct if and only if:
\begin{itemize}
    \item $(W, C)$ is a correct setup for $V^s$, and
    \item All $v \in V^d$ evaluate to true in all steps $s \in E$.
\end{itemize}

\end{definition}

Accordingly, an execution is not correct whenever either one of the static constraints is hurt or one of the dynamic \VC{}s in at least one step of the execution. We call a step $s \in E$ for which all $v \in V^d$ hold a \emph{correct step}; all other steps are called \emph{erroneous}. Naturally, the first erroneous step is of particular importance, as usually (but not necessarily) \DAW{} execution will stop at this point.

Our notion of \VC{}s clearly has limitations in terms of expressiveness, and we can envision several extensions. For instance, we only introduced \VC{}s that affect only single steps, single tasks, single dependencies, and single nodes. Thus, we have no notion for expressing constraints that, for instance, ensure (1) that two consecutive tasks in a workflow are scheduled on the same node (because we know of side effects not modelled in the \DAW{}), or (2) that the total size of all intermediate files may not exceed a certain threshold (because there is a quota on available disk space). We leave such types of \VC{}s for future work.

\section{Concrete Validity Constraints for \DAW{}s}
\label{sec:vcs}
After having defined \DAW{}s, their components, \VC{}s, and the formal relationships between \DAW{}s and \VC{}s in an abstract manner, we shall now introduce a broad collection of different concrete \VC{}s. We do not aim for completeness but for a representative set that illustrates the spectrum of functionalities that can be covered when using \VC{}s as first-class primitives for \DAW{} languages. Naturally, one can envision further constraints up to arbitrary user-defined \VC{}s, provided a proper specification language for them is defined. Note that none of the \VC{}s we discuss is completely new; instead, many of them can be found either implicitly or explicitly in other research fields, such as integrity constraints in databases or pre-/post conditions in programming languages; we shall discuss these related lines of research in \secref{sec:related-work}.

We shall present \VC{}s in three steps. We shall first list them grouped by the component of a \DAW{} system they address, namely swetup, task, or file, accompanied by an intuitive explanation and a classification into `static' or `dynamic'. In \secref{sec:vc-properties}, we distinguish several properties of \VC{}s to enable a more fine-grained distinction. These properties will allow to systematically characterize \VC{}s in \secref{sec:vc-characterization}.

Based on related ideas in other fields and own experience in \DAW{} development, we consider the following \VC{}s as particularly important. We group them according to the part of a \DAW{} / infrastructure they primarily affect. Note that not all of them have the same level of abstractions; in some cases, we rather describe a type of \VC{} than a concrete \VC{}. For instance, we introduce a general \VC{} for file properties instead of one distinct \VC{} for every such property.

\subsubsection*{Setup-related \VC{}s}
This set of \VC{}s are related to the particular combination of a \DAW{} and the cluster it should be executed on. Two of them are intended to be controlled before the \DAW{} execution starts and are thus static. The third is inherently dynamic.

\begin{description}
\item[resource availability:] The nodes within a cluster must fulfill certain requirements in terms of available resources, such as minimal main memory, minimal number of allocated \CPU{} hours, or availability of a \GPU{} of a certain type. This \VC{} could be defined with two different semantics: In  \emph{at-least one node}, at least one node of the cluster must fulfill the constraints; in \emph{all nodes}, all nodes must do so.
\item[file must exist:] File must exist and must be accessible. Thus may, for instance, affect certain reference or metadata files, but can also be used to ensure availability of input files of the \DAW{}. This \VC{} could also be defined either in \emph{at-least one node} or \emph{all nodes} semantics; however, the latter is more common.
\item[infrastructure health:] A node respond to request from the \DAW{} engine prior or during a \DAW{} execution. Such constraints are often implemented with the help of a heartbeat-style infrastructure.
\end{description}

\subsubsection*{Task-related \VC{}s}
Task-related \VC{}s describe properties of a concrete task of a \DAW{}. Many of them can be defined either statically or dynamically.

\begin{description}
\item[executable must exist:] During execution, any concrete task must be scheduled on some node in the cluster. The program executing this task must be available on this node. Can be defined statically, which requires that all nodes in the cluster maintain executables of all tasks in the \DAW{}, or dynamically, which allows for temporary installation (and subsequent deletion) of executables of tasks as part of their scheduling.
\item[resource availability:] Before starting a task on a given node, certain requirements in terms of available resources must be fulfilled, such as minimal main memory, minimal number of allocated \CPU{} hours, or availability of a \GPU{} of a certain type.
\item[configuration parameters:] Parameters for execution of a task within a \DAW{} must be valid, e.g., have a value within a certain range or of a certain format. Is typically defined dynamically as many arguments of tasks are created only at runtime, such as the names of input / output files.
\item[licence valid:] Some tasks might require a valid licence to start. Can be defined statically (test for general availability of a valid license for all tasks in a \DAW{}) or dynamically (test for concrete availability of a valid license as part of task scheduling). The latter is important then the number of possible concurrently running tasks is constrained by a volume contract.
\item[metamorphic relations:] The input/output-relationship of a task can be characterized by a reversible function. After executing a task, the concrete pair of input/output must have this relationship.
\item[tasks end within limits:] The runtime of a particular tasks can be constrained by a \VC{} on its maximal runtime. Such a constraint can help to identify stragglers.
\item[task ends correctly:] Execution of a particular task must end with a predefined state or output message.
\end{description}

\subsubsection*{File-related \VC{}s}
File-related \VC{}s control the management of files within the infrastructure. Using our definitions from \secref{sec:definitions}, this also encompasses dependencies and hence data exchange between tasks.

\begin{description}
\item[file must exist:] File must exist and must be accessible before starting a task on a node.
\item[file properties:] A file must fulfill certain criteria, such as file size, format, checksum over content, or creation time. Can be defined statically (properties of metadata files) or dynamically (properties of files generated during \DAW{} execution).
\item[folder exists:] Certain folders must exist and must be readable before/after execution of a \DAW{}/task.
\end{description}

\subsection{Properties of \VC{}s}
\label{sec:vc-properties}

We so-far classified \VC{}s only very broadly into two classes based on the point in time  when they can be checked in principle. However, there are many more dimensions by which \VC{}s can be characterized. For instance, violations of \VC{}s can have different levels of severity; while some must result in an immediate stop of the \DAW{} execution, such as in the case when a task in the \DAW{} requires an amount of main memory that none of the nodes of a cluster can provide, others might be interpreted rather as a warning, such as an improbable yet not impossible file size. Some \VC{}s must be checked before a task starts, such as the available resources on the node it is scheduled on, some after a task ends, such as its result status, and a third class of \VC{}s requires continuous control during task execution, for instance to ensure termination within a runtime limit.

\tabref{tab:VCKeyword} provides six different properties (or dimensions) by which \VC{}s can be characterized. These dimensions are mostly independent of each other and all have their own importance. For example, knowing whether a constraint is `hard' or `soft' is equivalent to knowing whether it expresses a mandatory requirement or not.
The `affected object' informs how to track the constraint and what might be affected if it is violated.
\enlargethispage{1.2ex}

\begin{table}[t!]
\caption{Dimensions by which Validity Constraints can be characterized.\label{tab:VCKeyword}}
{\footnotesize \setlength{\tabcolsep}{2pt}\renewcommand{\arraystretch}{1.2}

\begin{tabular}{>{\raggedright\arraybackslash}p{.29\textwidth}|p{.69\textwidth}}
\hfil\textbf{Dimension}                               & \hfil\textbf{Description and possible values}
\\ \hline
Severity            & Describes whether a \VC{} must be fulfilled or not; non-mandatory \VC{}s implement plausibility checks. Possible values: \emph{hard} implies immediate stop of \DAW{} execution; \emph{soft:} issues a warning, for instance in the \DAW{} log.
\\ \hline
Affected object     & Describes the type of object addressed by a \VC{}. This dimension was used to group \VC{} in the previous text. Possible values: \emph{setup}, \emph{task}, and \emph{file}.
\\ \hline
Type                & Describes the type of a \VC{}. Possible values: \emph{static}; \emph{dynamic}. See also \secref{sec:vc-def}.
\\ \hline
Time of check       & Describes the point in time when a \VC{} should be checked. Possible values: \emph{before, \before:} check before task starts on a given node; \emph{after, \after:} check after a task has finished; \emph{during, \during:} check periodically during task execution.
\\ \hline
Component           & Describes the component in the \DAW{} architecture (see \secref{sec:daw-infrastructure} and \figref{fig:daw-infrastructure-architecture}) which is responsible for controlling a \VC{}. Possible values: \emph{execution engine} \EE{}; \emph{scheduler} \Ss{}; \emph{resource manager} \RM{}; \emph{monitoring} \M{}. 
\\ \hline
Recoverable         & Describes whether the \DAW{} system can try to recover from the error automatically. Possible values: \emph{yes} ($+$), \emph{no} ($-$), \emph{maybe} ($\pm$).
\\ \hline
\end{tabular}}
\end{table}

Such a more fine-grained classification for \VC{}s enables differentiating techniques and therefore enables a common shared understanding and objective discussion about \VC{}s. Newly found \VC{}s can be contrasted and grouped with other validity constraints using a given classification. Classifications can also help identifying new \VC{}s, by looking for a \VC{} that fulfills a certain combination of properties.

\subsection{Formal Characterization of \VC{}s}
\label{sec:vc-characterization}

In this section, we introduce a classification for Validity Constraints for \DAW{}s. First, we will explain why a classification is helpful in this context. Then, we present the properties alias dimensions we deem relevant to classify \VC{}s and then classify the introduced \VC{}s in \secref{sec:vcs} accordingly.

Using the properties and dimensions of \tabref{tab:VCKeyword}, we classify the selected \VC{}s from~\secref{sec:vcs} in \tabref{tab:VCWF-properties} and
observe the following general trends and traits for the selected \VC{}s. Most \VC{}s are either hard constraints or can be both hard and soft constraints.
This characteristic is most likely because there is less value in a \VC{} that never indicates an error. There are two big groups regarding the time a \VC{} is checked: many \VC{}s are checkable either `before' or `during' the execution; few \VC{}s are checkable `after' the execution. This deviation is most likely caused by preconditions and invariants being more common than means to check postconditions. The most predominant component for checking \VC{}s is the \EE{}, which is tightly coupled to almost all of the dynamic \VC{}s, as the \EE{} is inherent to the execution. The time of check also correlates with the discreteness of the checks. If a \VC{}s' time of check is `before' or `after', it is usually `discrete'. If the time of check is `during', the \VC{} is usually `continuous'. As we do not have enough examples of `triggered' constraints, we cannot point out similar correlations for those. Many \VC{}s  are not limited to workflows but are also applicable in related fields, i.e., they represent more general constraints. Many constraint violations are `recoverable' as violating them can be caused by spurious problems.

\begin{table}[t!]
\caption{Validity Constraints for workflows.\label{tab:VCWF-properties}}
        {\footnotesize\setlength{\tabcolsep}{2pt}\renewcommand{\arraystretch}{1.3}
\begin{center}
\begin{tabular}{l|c|c|c|c|c|c}
  \hfil Constraint                                &
  \rotatebox{60}{h(ard)/s(oft)/b(oth)}            &
  \rotatebox{60}{affected object}                 &
  \rotatebox{60}{time of check}                   &
  \rotatebox{60}{responsible component}           &
  \rotatebox{60}{s(tatic)/d(ynamic)/p(osthoc)}    &
  \rotatebox{60}{recoverable?}\\ \hline
\hline 
\multicolumn{7}{l}{\emph{setup-related:}}\\
resource availability      & h & $P_C(c)$ or $P_T(s)$         & \before~\during & \Ss{},\M{}  & d   & $+$\\ \hline
file must exist            & h & $P_D^i(s)$, $P_D^o(s)$       & \before~\after  & \EE{}   & d/p    & $\pm$ \\ \hline
infrastructure health      & b & $P_C(c)$, $P_C(s,t)$         & \during         & \M{}    & d      & $+$\\ \hline
\hline 
\multicolumn{7}{l}{\emph{task-related:}}\\
  executable must exist      & h & $P_T(s)$ 
                                                  & \before         & \EE{}   & d   & $-$ \\ \hline
resource availability      & h & $P_C(c)$ or $P_T(s)$         & \before~\during & \Ss{},\M{}  & d   & $+$\\ \hline
configuration parameters   & b & $P_T(t)$                     & \before         & \EE{} & d   & $-$\\ \hline
licence valid              & h & $P_T(t)$                     & \before         & \EE{}   & d   & $\pm$\\ \hline
metamorphic relations      & b & $P_T(t)$                     & \after          & \EE{}  & d/p  & $\pm$\\ \hline
tasks end within limits    & h & $P_T(s)$                     & \during         & \Ss{},\EE{} & d/p & $\pm$\\ \hline
task ends correctly        & b & $P_T(s)$                     & \after          & \EE{}   & d    & $\pm$ \\ \hline
\hline 
\multicolumn{7}{l}{\emph{file-related:}}\\
file properties            & b & $P_D^i(s)$, $P_D^o(s)$     & \before~\after  & \EE{},\RM{} & d   & $\pm$\\ \hline
file must exist            & h & $P_D^i(s)$, $P_D^o(s)$     & \before~\after  & \EE{}   & d/p  & $\pm$ \\ \hline
folder exists              & h & $P_D^o(s)$                 & \before         & \EE{}   & d   & $\pm$\\ \hline
\hline
\end{tabular}
\end{center}}
\end{table}

\section{Validity Constraints in Related Fields}
\label{sec:related-work}

In the following, we first look at implicitly and explicitly defined \VC{}s one can find in research fields related to workflow technology. Technologies similar to \VC{}s are commonly used throughout different subjects, yet usually under different names and often not in an explicit manner. Many of the \VC{}s we detected have already been covered in \secref{sec:vcs}. Others do not apply to scientific workflows, are highly domain-specific, or go technically beyond the expressiveness of our underlying model. In \secref{sec:vc_workflows}, we discuss prior work in \VC{}-related concepts in scientific workflow research and conclude this section in~\secref{sec:vc_current} with a survey of validity checking mechanisms in selected current workflow languages or systems, namely \CWL{}, Nextflow, Snakemake, Airflow, Spark, and Flink.

\subsection{\VC{}s in Other Fields of Research}\label{sec:otherfields}
First, we look at \VC{}s in other fields of research.

\paragraph*{\emph{Database Management Systems:}}
Relational databases store data in the tabular format following the relational data model, essentially consisting of tables, attributes, and values of some basic data type. This data can be queried by declarative languages like \SQL~\cite{GM14}, which computes a result relation from algebraic expressions of selection, project, and join operations over base tables. Since the early days of the relational model, it was enriched by so-called integrity constraints (\IC{}s)~\cite{GA93}, which are logical formulas describing some property of the values of an attribute that must always be true. Typical systems differentiate between different types of constraints:
\begin{itemize}
    \item Constraints on individual values, such as value range constraints, enumerations of allowed values, or `not null' constraints.
    \item Constraints referring to all values of an attribute, especially the `unique' constraint demanding that all values of an attribute are pair-wise different.
    \item Constraints that relate values of one attribute to values of another attribute. The most prominent constraint of such kind is the `foreign key' constraint, demanding that every value of a dependent attribute is also present as value of another attribute in a different table.
\end{itemize}
On top of these, many database engines also allow the programming of user-defined constraints by using `trigger' operations, which are executed upon specific actions performed on some value in the database, such as insertion or deletion of a tuple.

\IC{}s for databases apparently are highly similar to \VC{}s for workflows. However, there are also important differences. The most important one is that integrity constraints are defined over a given and persistent database. Once defined, they are enforced on each change request of the data in the database. As a consequence, the change is accepted or refuted. Databases thus can never be in an inconsistent state regarding a defined set of active \IC{}s. In contrast, \VC{}s for workflows are defined over a transient process of computations and must be controlled alongside the \DAW{}'s execution progress. Some \VC{}s aim to prevent inconsistent states, especially those checking the pre-conditions of a task, while some can only react after an inconsistent state occurs, such as post-conditions of tasks. Furthermore, current database systems are monolithic systems that incorporate \IC{} control in their code base at defined places, while \DAW{}s are executed over infrastructures composed of multiple independent components, which makes implementation of a \VC{} control machinery more challenging (see \secref{sec:vc-impl}).

\paragraph*{\emph{Model Checking}} is a technique for automatic formal verification of finite state systems. The model checking process can be divided into three main tasks~\cite{Clarke2018}:
\begin{itemize}
    \item \emph{Modeling:} Convert a design (software, hardware, \DAW{}) into a formalism accepted by a model-checking tool.
    \item \emph{Specification:} State the properties a design must satisfy (i.e., some logical formalism, such as modal or temporal logics).
    \item \emph{Verification:} Check if the model satisfies the specification (ideally completely automatic).
\end{itemize}
Especially the second task (specification) and the first task (modeling) are related to \VC{}s. During modeling, the \DAW{} is translated into a Kripke transition system~\cite[Ch.~3]{Clarke2018}, an automaton with states (tasks) and transitions. Each valid path in the Kripke transition represents a valid execution path in the \DAW{}, thus ensuring that the necessary task execution order is respected. Temporal logic~\cite[Ch.~2]{Clarke2018} is used, for example, to define \VC{}s locally or globally in the specification. In other words, the constraints can be on the state (task) level, which can indicate that there exists a task along the path that fulfills a given condition or constraint. Furthermore, constraints can also be defined on the path-level, meaning that there exists a path generated from a state that holds true for a given condition~\cite{Beyer2018}.

\paragraph*{\emph{Business Process Management (BPM)}} studies workflow processes in business-related areas to improve business process performance. There are different relevant perspectives to consider. The \emph{control-flow perspective} models the ordering of activities and is often the backbone of BPM models. Organizational units, roles, authorizations, IT systems, and equipment are summarized and defined in the \emph{resource perspective}. Furthermore, the \emph{data or artifact perspective} deals with modeling decisions, data creation, forms, etc. The \emph{time perspective} addresses task durations but also takes fixed task deadlines into account. Lastly, the \emph{function perspective} describes activities and related  applications~\cite{van2016business}.
Process models can use conditional events to define business rules. A conditional event allows a process instance to start or progress only when the corresponding business rule evaluates to true. When handling exceptions in BPM, validity constraints can be internal (caused inside the task) or external (caused by an external event) exceptions. Another constraint is the activity timeout, where an activity exceeds the predefined time for execution~\cite{Dumas2013}.

\paragraph*{\emph{Software Engineering and Programming Languages:}}
Software engineering is concerned with the design, implementation, and maintenance of software systems. Data types are one of the tools to help in this process, as type specifications define the interface of how different components in a system are allowed to interact. Using inappropriate data types will lead to unspecified and likely invalid behavior. Thus, enforcing type constraints of the processed data is critical for ensuring the validity of software systems.

Fortunately, most programming languages these days support a kind of type checking. Dynamically typed languages such as Python do this at runtime, whereas statically typed ones such as Java check the type constraints already at compile time. If a type constraint is violated, an appropriate error message is usually returned, so the developer knows there is a conflict of mismatching types.

Assertions and exceptions are other constructs provided by many programming languages to check user-defined validity constraints at runtime, for example, to check for the existence of a particular file. Assertions are also commonly used to specify the correct behavior in a software test. Bertrand Meyer~\cite{meyer1988eiffel} designed the Eiffel programming language to increase the quality and reliability of software systems. Meyer coined the term Design-by-Contract, a central concept of the Eiffel language, which is a methodology to design correct and reliable systems using assertions,  preconditions, postconditions, and class invariants. Design-by-Contract has its roots in Hoare Logic~\cite{pratt1976semantical} for proving program correctness.

Wasserman and Blum~\cite{wasserman1997software} proposed result checking, where a system has a separate program whose only purpose is to check the correctness of the results. The difference from software testing is that the checker needs to satisfy stringent properties regarding its reliability and running time.

The Rust programming language\footnote{\url{https://www.rust-lang.org/}}~\cite{rustbook} recently emerged as a language designed with safety in mind, especially for concurrent programs. Rust programs can be guaranteed memory safe by using the borrow checker that validates references to memory. Through its ownership system, Rust controls who has what access to memory locations, avoiding situations where multiple threads have mutable access to the same variable concurrently.

\paragraph*{\emph{Service Composition and Interface Constraints:}}
\label{sec:composition}
The topic of automatic service composition is also the main focus of the book by Tan and Zhou~\cite{2013-tan}. It discusses, for example, the verification of service-based workflows, quality-of-service (QoS) aspects, deadlock detection, and dead path elimination. Based on interface descriptions they verify the automatic composability of workflows. As a foundation to analyze properties (e.g., deadlock detection) of \DAW{}s Petri nets~\cite{diaz09}, $\pi$-calculus~\cite{milner99}, process algebra~\cite{Fokkink00}, or automata (linear temporal logic)~\cite{Baier08} can be used. When services or tasks are not directly composable, one can look for mediator tasks that make interfaces compatible. In terms of validity constraints, the work mainly focuses on the validity of interfaces and data formats between tasks, which may be overcome with mediators.

\paragraph*{\emph{Machine Learning Operations (MLOps)}}
encompasses the end-to-end process of employing machine learning models and pipelines in application contexts. For instance, a video streaming service may utilize a machine learning model to recommend content to users or to place relevant advertisements on its platform. Typically, the model is embedded in an end-to-end pipeline designed, developed, and continually integrated into application contexts. This integration involves reoccurring tasks, such as the curation, filtering, and preprocessing of datasets, plus the design, training, and validation of models~\cite{2017TFX}.

Validity constraints for models in the context of MLOps address the robustness and prediction quality of models. While it is usually hard to comprehensively model correctness throughout entire machine learning pipelines, error cases can be anticipated and checked in the pipeline input and output. For instance,
\begin{itemize}
    \item Quality metrics may cross-check a pipeline output to detect performance deterioration.
    \item Checks for changes in the distributions of input data features may detect drifts that could lead to a decreased predictive performance. More specifically, a feature may be assumed to always lie in a particular value range or have a specific value distribution.
    \item Checking that infrastructure requirements are satisfied (e.g., available \GPU{} memory to load and execute a model checkpoint) can prevent exceptions or application crashes.
\end{itemize}

Machine learning pipelines are often designed for diverse infrastructures and complex, custom technology stacks.
While research on platforms to support the MLOps process exists (e.g., TFX~\cite{2017TFX}, MLFlow~\cite{zaharia2018mlflow}, or Kubeflow~\cite{bisong2019kubeflow}), there exists no widely accepted standard platform for end-to-end machine learning operations.
Also, automated and comprehensive validity checking is not part of their functionality. Yet, pipelines undergo continuous changes, driven by trends in real-world input data, conformance regulations, AI fairness issues, or AI safety concerns. Each change has the potential to introduce errors to the pipeline. Hence, the need for standardized quality assurance solutions in the MLOps process is apparent.

\subsection{Previous Work on \VC{}s for Scientific Workflows}\label{sec:vc_workflows}

Scientific Workflows are \DAW{}s in the scientific data analysis domain (\secref{sec:intro}). Typically, they build on a Scientific Workflow Management System, which encompass workflow languages, execution engines, a form of resource management, and a form of data exchange; the later two components are often delegated to infrastructure components like shared file systems or resource managers. Over the years, many such systems were developed with differing features and capabilities~\cite{DGST09}. Validity constraints---although they are a vital ingredient for portability, adaptability, and dependability as discussed in \secref{sec:intro}---often remain implicit and unchecked in these systems~\cite{CBB+17}. Some research addresses only very specialized aspects, such as Rynge et al. who focus solely on detecting low-level data corruption (as hard \VC{}s), for instance, after caused by network or hardware errors~\cite{RVD+19}. In the following, we discuss some prominent systems or perspectives from the viewpoint of validity constraints.

\paragraph*{Semantic Workflows:}
\label{sec:semantic}
Semantic workflows denotes a class of workflow languages that build on an elaborated, often domain-specific type system or ontology~\cite{GSVHRGMSK11}. With this ontology, data that is to be exchanged between tasks are assigned a specialized type (such as ``genomic reads from machine X'' instead of the basic ``set of strings''), tasks are assigned a type (such as ``read mapper for genomic reads''), and the IO channels of tasks are assigned types. Types are arranged in a specialization hierarchy which allows inference regarding type compatibility or workflow planning~\cite{LPI+21}. For instance, Lamprecht introduced a workflow language that allows for the definition of semantic constraints, leading to methodologies for model-guarded and model-driven workflow design~\cite{Lam13}. Another example is the semantics-based `Wings' approach to workflow development and workflow planning~\cite{GRKGGMD11}.

Types, i.e., semantically defined concepts, together with compatibility checking are a form of validity constraints. The are usually defined statically and can be checked before workflow execution based on annotation of the workflow components. They operate on a different level than the \VC{}s we defined. However, such systems are (yet) rarely used in practise because they require all data files and all tasks to be used in a workflow to be annotated with concepts from a consistent ontology. In quickly changing fields like scientific research where \DAW{}s are often explorative, this requirement makes development cumbersome and inflexible. It also requires significant effort in community-driven ontology design and maintenance~\cite{IKJBUMMLPR13}.

\paragraph*{\emph{\AWDL{} Based Workflows and Data Constraints:}}
\label{sec:AWDL}

Qin and Fahringer~\cite{2012-qin-fahringer} use the abstract workflow description language (\AWDL{}), an \XML{}-based language expressing workflows. \AWDL{} allows describing the directed acyclic graph (\DAG{}) of tasks with their conditions and execution properties (parallel, sequential, alternative paths), etc. Further, it allows specifying constraints for the runtime environment, optimization, and execution steering. The approach follows a \UML{}-based workflow modeling and modularization.  The specification of data representations and activity types with ontologies aims for automatic semantic workflow composition and automatic data conversion.

\AWDL{} supports simple properties and constraints per data port and task, such as read-only or read-write data, expected output size, memory usage, required \CPU{} architecture, etc. It also supports constraints on the data distribution like ``a task only needs the first index'', ``a task can work on single items'', ``a task needs a window of $x$ items'', etc. The typed data sources and sinks help the automatic composability of workflows and necessary data conversion tasks.

\paragraph*{\emph{Temporal Constraints:}}
\label{sec:temporal}
\begin{table}[t!]
  \caption{Overview on the support of temporal QoS
    constraints~\cite{2012-liu}\label{tab:temporal}} 
  \centering\includegraphics[width=.8\linewidth]{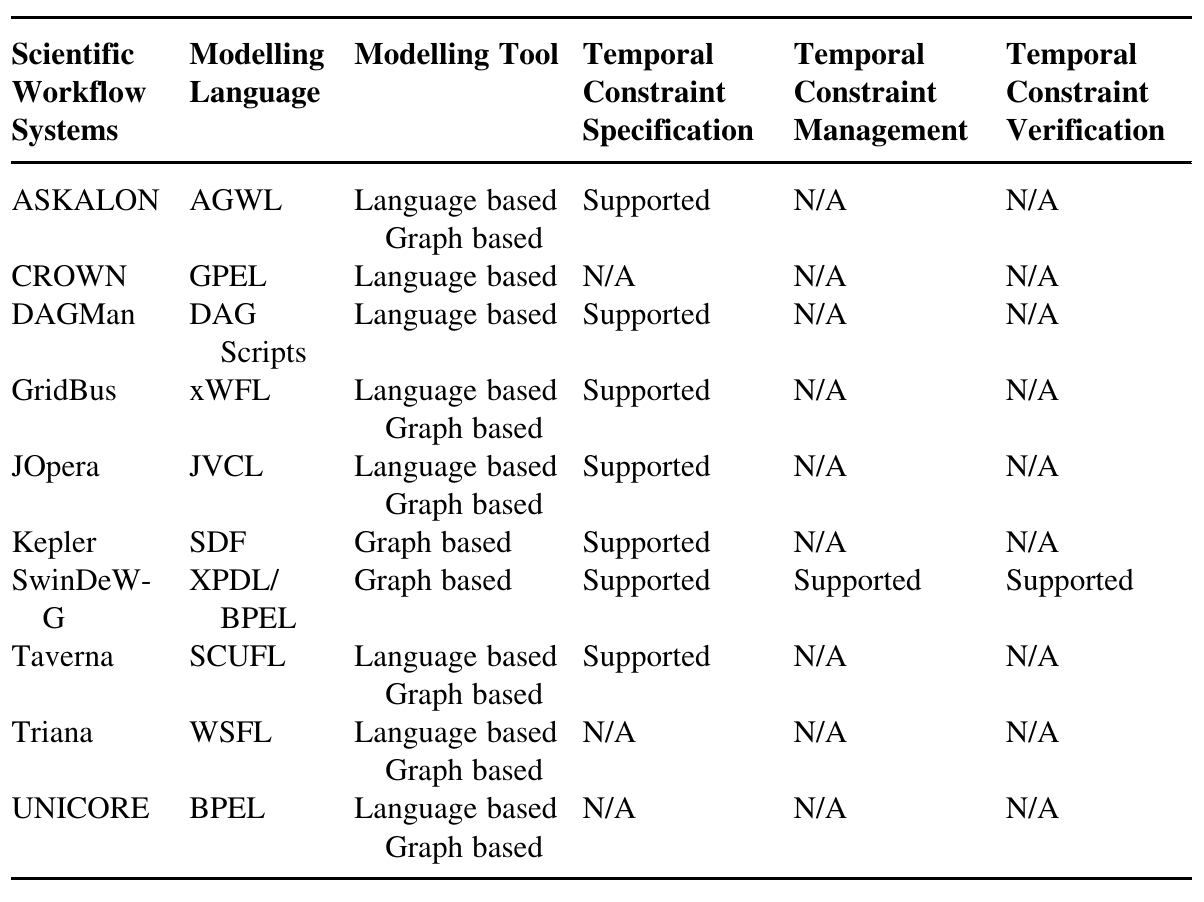}
\end{table}
Liu, Yang, and Chen discuss temporal constraints in scientific workflow systems~\cite{2012-liu}.  They argue that fixed time constraints are often too strict, and their violation not necessarily indicates a failing (or otherwise wrong) workflow execution.  Instead, they introduce the concept of probabilistic temporal constraints, e.g., 90\,\% of tasks of class `A’ finish within 60 minutes. They distinguish the components of \emph{setting} temporal constraints, \emph{monitoring} temporal consistency, and \emph{handling} temporal violations. Checkpoints can be used for re-execution and temporal checks.  To overcome constraint violations they distinguish \emph{statistically recoverable} temporal violations and \emph{statistically non-recoverable} temporal violations.  The former can be handled by doing nothing, or re-scheduling, and the latter by adding resources, stopping and restarting tasks or workflows, or workflow restructuring.  \tabref{tab:temporal} (copied from~\cite[p. 57]{2012-liu}) lists several scientific workflow systems and their capabilities to handle temporal constraints.  In essence, (statistical) temporal constraints can be defined for \mbox{(sub-)\DAW{}s} to check the validity of executions.

\paragraph*{\emph{Provenance and Reproducibility Constraints:}}
\label{sec:provenance}
In their high-level vision paper~\cite{2018-deelman}, Deelman et al. state (emphasis added): ``During and after the workflow execution, the capture of provenance information and its use \emph{to validate performance and correctness} and to \emph{support data reuse and reproducibility} are areas where \emph{much research is needed.''} The overview paper identifies a bunch of relevant challenges. Unfortunately, concrete solutions are still missing for most of them. The mentioned provenance data could help to check reproducibility and predictable performance, for example. Another important aspect the paper discusses is the accuracy and similarity of results to guarantee scientific reproducibility. A \DAW{}'s re-execution in another execution environment raises the question: When are result deviations significant and become unacceptable? Typically, we cannot expect byte-level equivalence of results, especially when dealing with floating-point arithmetic. Although, the results for the same inputs should resemble each other to a certain degree for sound workflows.

Related to the issue of reproducibility is the topic of consistency checking for provenance. For instance, the \PROV{} model for provenance defines several constraints on concrete instances of the model~\cite{2013Moreau}. Specifically, four types of constraints are defined: uniqueness constraints, event ordering constraints, impossibility constraints, and type constraints. From a workflow perspective, especially event ordering constraints are relevant, which are similar to our notion of (valid) executions introduced in \secref{sec:daw-infrastructure}. However, \PROV{} is a model for storing provenance information and not for executing workflows. Actually, a workflow executed by a valid schedule should, be definition, always produce provenance information with valid event ordering. \PROV{}-Wf~\cite{CSOOODM13} and prov\ONE{}~\cite{MDBCL13} are two different extensions to \PROV{}, compared in~\cite{OM+18}, that can also model workflows directly (called prospective provenance in this community), but neither elaborats on validity constraints.

\subsection{\VC{}s in Current Workflow Systems}\label{sec:vc_current}
After this overview of validity constraints of different fields, we next describe the state-of-the-art in validity constraint definition and checking in actual systems. To this end, we look at a selection of current popular state-of-the-art workflow systems and examine if and how they support the application of validity constraints.

\paragraph*{\emph{Common Workflow Language (\CWL{})}}
is an open standard that facilitates the description of command-line tool execution and workflow creation. It is still under active development~\cite{Amstutz2016}. The ways to define validity constraints are currently limited but subject to extension. So far, \CWL{} supports a dynamic definition of resource requirements enabling the optimization of task scheduling and resource usage without manual intervention. Additionally, it allows the specification of software requirements. Both the resource and software requirements are expressed as hints. Workflow engines may consider or ignore these annotations as \CWL{} is merely a workflow language and standard but does not provide a full-fledged execution engine other than a simple proof-of-concept runner. For better validation of workflow connections, it is recommended to use file format identifiers~\cite{crusoe2022methods}. An extension currently under discussion is the addition of input value restrictions.\footnote{\url{https://github.com/common-workflow-language/common-workflow-language/issues/764}}

\paragraph*{\emph{Nextflow}}
is a workflow system that provides its own domain-specific language to compose user-provided tasks into workflows~\cite{DiTommaso2017}. Although it is mainly used in the bioinformatics domain, Nextflow can be used to build workflows in any domain. Recently, the Nextflow developer team introduced their new language `{\smaller DSL2}' to build Nextflow workflows. They point out that the next focus is to take advantage of the improved modularization capabilities of {\smaller DSL2} to support the testing and validation of process modules. We are not aware of any built-in functionality to define or check validity constraints of the workflows currently, though. Note that in \secref{sec:vc-impl} we will describe a prototype implementation of \VC{} for Nextflow.

\paragraph*{\emph{Snakemake}} is a workflow management system that uses a Python-based language to define and execute workflows. Each rule in Snakemake specifies input and output files, along with any parameters or commands needed to produce the output from the input. The rules can be chained together to form a directed acyclic graph that represents the dependencies between the rules~\cite{KosterR18}. While Snakemake ensures that each rule is well-defined and the workflow is reproducible, it does not, as far as we know, provide a formal mechanism for specifying validity constraints or checking the correctness of the workflow at runtime. Although, Snakemake supports a dry-run with the command line option `\texttt{-n}' that can be used to check whether the workflow is defined properly and can also provide a rough estimate of the required computational time necessary to execute it. Furthermore, Snakemake checks for the existence of a task's defined output files after its execution. For further checks, such as checking for them to be non-empty, users are advised to implement that by shell commands manually to provoke a non-zero exit status of the task.\footnote{\href{https://snakemake.readthedocs.io/en/v7.25.0/project_info/faq.html\#how-do-i-make-my-rule-fail-if-an-output-file-is-empty}%
{\texttt{https://snakemake.readthedocs.io/en/v7.25.0/project\_info/faq.html}}}

\paragraph*{\emph{Apache Airflow}}
is a workflow management system created in 2014 by Airbnb~\cite{AirFlow}. Workflows in Airflow are created using the Python \API{}. Airflow does not explicitly provide functionality targeted at checking the validity of workflows. Instead, they provide a best practices section in their documentation with a description of testing of airflow workflows. In this description, the authors suggest manually inserting customized checks into the workflow to ensure results are as expected. However, such a check is simply another user-defined task inside the workflow, and there are no specific airflow constructs to help build such checks or to react when checks fails.

\paragraph*{\emph{Apache Spark,}} started in 2009 at UC Berkeley, is a workflow engine for large-scale data analysis~\cite{ZXWDADMRV16}. Spark workflows are defined via \API{}s in Java, Scala, Python, or R. Apache Spark does not seem to support validity constraints for their workflows. Therefore, users need to come up with their own validation schemes.

\paragraph*{\emph{Apache Flink}}
is a data analytics engine unifying batch and stream processing~\cite{CKEMHT15}. Akin to Spark, Apache Flink workflows are created using Java, Scala, or Python \API{}s. In a document for the nightly build of Apache v1.15, the Apache Flink team introduces a new non-stable minimum viable product named ``Fine-Grained Resource Management''. This new feature will allow workflow developers to specify the resource requirements manually for each task. While this feature's primary objective is to improve resource utilization, this may provide the possibility for resource-based validity constraints. Aside from that, Flink offers extensive support for local testing and validating workflows with constructs such as test harnesses and mini clusters.

\section{Implementing \VC{}s for \DAW{}s as Contracts in Nextflow}
\label{sec:vc-impl}

We implemented a prototype of \VC{}s in the popular workflow system Nextflow to validate our conceptual model in practice. Details of the implementation and its evaluation are beyond the scope of this paper and will be published elsewhere. In brief, we added two new directives called `\texttt{require}' and `\texttt{promise}' into the Nextflow specification language, which allow us to insert code for validity constraints into task definitions. By incorporating \VC{}s into the workflow definition language, we can leverage the existing language tools available in Nextflow, such as Groovy. This approach eliminates the need for a new \VC{} language with a separate syntax and interpretation infrastructure. As a result, \VC{} writers benefit from the familiar language tools and do not have to learn a new specification language~\cite{fahndrich2010embedded}. The two newly added primitives are part of an extension to the \DAW{} model borrowing the concept of contracts from software engineering, described in \secref{sec:otherfields}.

This contract-based approach allows adding a contract to each task in a workflow. Such contracts manifest as sets of requirements and promises checked immediately before (\before{}) and after (\after{}) the task execution to ensure that the task runs in the appropriate environment and produces valid results.
The new primitives `\texttt{require}' and `\texttt{promise}' that we added to the workflow definition language in Nextflow, enables the insertion of code for these contracts directly into the task definition. These contracts are then executed alongside the tasks on the cluster as arbitrary bash scripts, thanks to Nextflow's nature of compiling each task into a bash command script. To facilitate the creation of these contracts, we introduced auxiliary constructs with an internal domain-specific language (\DSL{})~\cite{fowler2005} to Nextflow's workflow definition language.
 The following example shows how to define \VC{}s in the form of a contract for a Nextflow process:

\noindent\begin{minipage}{\linewidth} 
\begin{lstlisting}
process x {
    input:
    [...]

    require([
        FOR_ALL("f", ITER("*.fa"),    // for all FASTA files (extension .fa)
           { f -> IF_THEN(
               COND("grep -Ev '^[>ACTGUN;]' $f"), // check lines' first char
               "exit 1")      // exit if first char is not one of ">ACTGUN;"
           }
        )
    ])

    promise([
        COMMAND_LOGGED_NO_ERROR(),     // auxiliary function checking stderr
        INPUTS_NOT_CHANGED()           // auxiliary function checking inputs
    ])

    """
    DATA PROCESSING CODE
    """
}
\end{lstlisting}
\end{minipage}

When Nextflow generates the command bash script for a process, it now places the code from the require block before the process code and the code from the promise block after the process code. The resulting program can be sent for execution on the cluster nodes. In this specific example, the process requires that all FASTA\footnote{\url{https://blast.ncbi.nlm.nih.gov/doc/blast-topics/}} files should have lines that start with certain characters, such as $\{>, A, C, T, G, U, N, ;\}$. This is a simple check to verify the file format before data processing. After processing the data, the task ensures that the process execution command did not encounter any errors and that the input files remain unmodified. These contracts are categorized as \emph{dynamic validity constraints} because they are code that runs alongside the task they are defined in. In terms of categorizing \VC{}s as described in \secref{sec:vcs}, these implemented contracts belong to the task-related \VC{}s. The task contracts can dynamically check node- and file-level properties, such as verifying that the current node has sufficient resources. However, they cannot check properties for all nodes.

We enhanced various real-world \DAW{}s in the field of bioinformatics with contracts to test the effectiveness and comprehensiveness of our contract-based approach to implementing validity constraints. This allowed us to identify common problems that arise during their execution and demonstrated how the specific notifications provided by broken contracts aid in debugging the \DAW{}s. Our investigation focused on three main areas: (1) the impact of runtime overhead on each task, (2) the amount of computation time that could be saved by aborting the \DAW{} early, and (3) how contracts enhance issue localization and explanation. Our experiments confirmed  that the specification even of simple contracts are very effective in supporting the identification of issues in real-world \DAW{}s, that they can save substantial compute time due to early aborts, and that the runtime overhead often is negligible, depending on the type of checks performed.

\section{Conclusions}
\label{sec:conclusion}

In this article, we introduced \VC{}s as a means to make implicit assumptions in data analysis workflows explicit, allowing a workflow engine automatically check their status and take proper action if needed. We defined a formal model connecting \VC{}s to the core elements of \DAW{}s, namely tasks for computations, files for data exchanges, and nodes for execution. Based on this formal model, we introduced different types of \VC{}s and classified them according to six dimensions. We extensively discussed similar concepts in various fields of research to show that (a) \VC{}s indeed are a vital and ubiquitous concept, but at the same time, (b) a unifying theory was missing, and (c) support for \VC{}s should be considered as partial at best in production or research systems. We hope our work will help to improve this situation by making \VC{}s an integral part of future \DAW{} languages and systems. \VC{}s can support debugging, save energy and time by an early failure of workflow executions, provide traceable warnings or error messages, and raise confidence in analysis results as they help making \DAW{}s more reliable.

Several extensions to our work are possible. In certain situations, e.g., IoT, \DAW{}s are increasingly often used to analyze data streams, which would pose specific requirements to validity checking and require fundamentally changing their semantics; for instance, the notion of failure would need to be revisited. One could also increase the expressiveness of \VC{}s by allowing constraints that affect groups of tasks (e.g., the total memory of a group of tasks scheduled on a node may not exceed the overall memory of the node) or groups of files (e.g., the files sent to different downstream tasks must be identical). \VC{} checking could directly link to counter actions; for instance, breaking a constraint about necessary memory on a node could result in feedback to the scheduler and trigger a re-scheduling of affected tasks. We leave such ideas for future work.

\section*{Acknowledgements}
This work was supported by the German Research Foundation (DFG) as CRC 1404, project 414984028.

\bibliographystyle{plain}
\bibliography{references}

\end{document}